\pdfoutput=1

\documentclass[conference,a4paper]{IEEEtran}

\usepackage[utf8]{inputenc}
\usepackage[T1]{fontenc}
\usepackage{lmodern}
\usepackage{microtype}
\usepackage[english]{babel}
\usepackage{xcolor}
\usepackage{amsmath}
\usepackage{amssymb}
\usepackage{graphicx}
\usepackage{adjustbox}
\usepackage{booktabs}
\usepackage{array}
\usepackage{tabularx}
\usepackage{ragged2e}
\usepackage{xurl}
\usepackage[hidelinks]{hyperref}
\usepackage{placeins}
\usepackage{tikz}
\usetikzlibrary{positioning,arrows.meta,fit,backgrounds,calc}

\newcolumntype{L}[1]{>{\RaggedRight\arraybackslash}p{#1}}
\newcolumntype{Y}{>{\RaggedRight\arraybackslash}X}
\renewcommand{\arraystretch}{1.13}
\definecolor{tablehead}{gray}{0.94}
\definecolor{tablegroup}{gray}{0.97}
\definecolor{envfill}{gray}{0.95}
\begin{document}

\title{LLM Non-Determinism in Deterministic Processes\\
{\large A Controlled Variation Space for Constraining the Propagation of Non-Determinism}}

\author{\IEEEauthorblockN{Paul Darius Mandl\IEEEauthorrefmark{1}, Peter Mandl\IEEEauthorrefmark{2}, Martin Häusl\IEEEauthorrefmark{2}}
\IEEEauthorblockA{\IEEEauthorrefmark{1}Findustrial GmbH, Schörfling am Attersee, Austria\\paul.mandl@findustrial.io}
\IEEEauthorblockA{\IEEEauthorrefmark{2}Munich University of Applied Sciences, Munich, Germany\\\{peter.mandl, martin.haeusl\}@hm.edu}}
\maketitle

\begin{abstract}
Large Language Models are increasingly used in software development processes
that rely on defined states, rules, and approvals. This creates a tension
between the flexibility of LLMs and the need for reliable and traceable
process states. We introduce \emph{Controlled Non-Determinism}, a
process-oriented approach that allows variation without letting it directly
determine the authoritative process state. We place LLM-based processing
within a controlled variation space that is bounded by a \emph{Deterministic
Envelope}. Before execution, we define the process-relevant properties of the
result and the exit conditions. We also record the task configuration. These
elements remain fixed during execution. At the end of the variation space, an
exit check determines whether the result satisfies the predefined conditions
and may become the next authoritative state. Different runs may produce
different admissible results while the transition to an authoritative state
remains controlled. We illustrate the approach with regulated software
development and derive design guidelines for LLM-based changes whose results
can be checked before they become part of the authoritative software state.
\end{abstract}

\vspace{0.2em}

\section{Problem Statement}
\label{sec:problem}

LLMs increasingly perform tasks in software development that can only be
described to a limited extent by conventional rules. Coding agents interpret
requirements, inspect repositories, generate implementation proposals, select
tools, modify files, and add dependencies or tests. We are interested in
settings in which this flexibility is useful but cannot simply be allowed to
determine the software state on its own. Regulated development processes still
require changes to be grounded in documented inputs, defined rules, checks,
and traceable approvals.

As long as an LLM output remains a proposal, variation may have no immediate effect on the software development process. This changes once the output influences subsequent actions. For example, an alternative implementation may affect interfaces or dependencies, while a different tool choice may modify external resources. Such variations can propagate through subsequent development steps and ultimately affect an authoritative software state.

A deterministic processing step can be represented in simplified form as
\begin{equation}
s_{k+1}=f_k(s_k,x_k)
\label{eq:det}
\end{equation}
where the same state $s_k$ and the same input $x_k$ are expected to produce
the same successor state. This assumption cannot be transferred directly to
LLM-based processing. In conventional LLM serving, even under inference settings intended to be deterministic, numerical effects, the dynamic grouping of requests into compute batches (dynamic batching), and other execution conditions may lead to different results \cite{Yuan2025Numerical,Gond2026LLM42,Zhao2026CoRun}.

For our purpose, asking whether two runs produce identical model outputs is
therefore the wrong starting point. We instead ask which differences survive
into downstream processing. Two different implementations may satisfy the same requirement and expose the
same relevant behavior even when their code structure differs.
Conversely, an initially small deviation may change the information state,
the sequence of tool calls, or ultimately the authoritative process state.

We therefore do not formulate the design problem as the suppression of all
variation. We ask how far variation may extend, where its effects must be
constrained, and under which conditions a varying result may cross into an
authoritative process state. The basis for that transition must remain
traceable as well.

To address this problem, we introduce the concept of \emph{Controlled
Non-Determinism}. It denotes processing in which non-deterministic variation is
deliberately permitted within a defined process region while its effects on
authoritative process states remain controlled. The associated
\emph{Deterministic Envelope} bounds this variation space between controlled
process states. Within the Envelope, LLM calls, tool interactions, and
internal processing steps may produce different execution paths and
intermediate results. Such variation is permissible as long as it remains
within the bounded region and does not enter an authoritative process state
without being checked. Only a defined exit check decides whether the produced
result satisfies the predefined conditions for adoption by the downstream
process.

The objective is therefore not deterministic repetition of the internal
execution, but the controlled use of non-deterministic processing while
constraining its effects on authoritative process states.

We make three contributions. First, we introduce a process model of Controlled
Non-Determinism that defines a controlled variation space, a notion of
process-relevant equivalence, an exit check, and a run artifact for traceability
(Section~\ref{sec:envelope}). Second, we derive a criterion for placing control
boundaries based on the potential process effect of variation and apply it to
regulated software development (Section~\ref{sec:applications}). Third, we
derive design guidelines for posing LLM tasks so that their results can be
checked and use the model to compare how current agent environments place
their control boundaries (Section~\ref{sec:fqanswers}).

\section{Related Work}
\label{sec:related-work}

Research on non-determinism in LLM-based systems addresses the emergence of
variation during inference, its propagation in multi-step systems,
architectural mechanisms for constraining it, and the traceability of the
resulting execution. For controlled processes, the key question is whether and
how internal variation affects authoritative process states.

Table~\ref{tab:related-work} provides an overview of work directly related to
the LLM-specific problem considered here. The classification follows the four
research strands discussed below. In addition to the primary focus of each
work, the table summarizes its contribution to the problem of integrating LLM
variation into deterministic processes in a controlled manner.

\begin{table*}[t]
\centering
\caption{Classification of selected recent work on non-determinism in LLM-based systems.}
\label{tab:related-work}
\scriptsize
\renewcommand{\arraystretch}{1.18}

\begin{tabularx}{\textwidth}{
@{}
>{\raggedright\arraybackslash}p{2.75cm}
>{\raggedright\arraybackslash}p{2.45cm}
>{\raggedright\arraybackslash}X
>{\raggedright\arraybackslash}X
@{}
}
\toprule
\textbf{Work} &
\textbf{Focus} &
\textbf{Main finding} &
\textbf{Relevance to this work} \\
\midrule

Atıl et al. \cite{Atil2025NonDeterminism},
Yuan et al. \cite{Yuan2025Numerical} &
Inference &
LLMs configured for deterministic behavior can still produce varying outputs
across repeated executions. Numerical effects are a documented cause. &
Non-determinism can arise within inference itself and must be distinguished
from deterministic decoding. \\

Gond et al. \cite{Gond2026LLM42},
S.~Zhao et al. \cite{Zhao2026CoRun} &
Deterministic inference &
Verification and padding techniques address batch- and kernel-induced causes
of variation. &
These works target the emergence of variation, whereas this work considers
its effects on the process. \\

Q.~Zhao et al. \cite{Zhao2025SAUP},
Xia et al. \cite{Xia2026Propagation} &
Propagation &
Uncertainty can be propagated and transformed across multiple steps of an
LLM-based system. &
Provides a basis for considering the reach of local variation. \\

Bjarnason et al. \cite{Bjarnason2026Randomness},
Yagubyan \cite{Yagubyan2026Consistency} &
Trajectories &
Repeated agent runs can develop different execution paths at an early stage. &
Different trajectories are unproblematic as long as they do not produce a
process-relevant difference in the successor state. \\

Mazhar et al. \cite{Mazhar2026Contamination},
Xie et al. \cite{Xie2026Cascades},
Singh and Pawar \cite{Singh2026Snowball} &
Error cascades &
Local errors and information deviations can affect subsequent states and
decisions. &
Motivates the need to constrain the effects of local deviations within the
process. \\

Srinivasan \cite{Srinivasan2026SDB} &
Architectural boundary &
The SDB defines the transition from an LLM proposal to a system action as a
contract consisting of Proposer, Verifier, Commit, and Reject. &
An SDB can realize an exit boundary but does not determine the extent of the
preceding variable region. \\

Strickland et al. \cite{Strickland2026Schema} &
Schema gating &
Conversational authority and execution authority are separated. Execution
occurs only after schema validation. &
Schema gates can be used as an exit condition of a variation space. \\

OWASP Agent Control Standard (ACS)
\cite{OWASP2026ACS,OWASP2026ACSRepo} &
Runtime control &
Standardized lifecycle hooks connect an observed agent to a Guardian that can
allow, deny, modify, escalate, or defer actions at runtime. &
ACS provides concrete technical intervention points for control boundaries,
but does not determine where those boundaries should be placed within the
process. \\

Kravchenko et al. \cite{Kravchenko2026APPA} &
Information-flow control &
APPA checks tool calls before execution and validates their realized returns
before admission to the agent context. Temporary child branches can locally
confine untrusted data. &
APPA realizes deterministic control points for information flows and external
effects. The Deterministic Envelope is more general because it also determines
the process-level placement and extent of a variation space. \\

Trooskens et al. \cite{Trooskens2026Compiled} &
Compiled AI &
LLM use is shifted into an upstream compilation phase. Validated artifacts are
then executed deterministically. &
Represents the limiting case in which the variation space ends entirely
before the actual runtime. \\

Qiu et al. \cite{Qiu2025Blueprint},
Kim and Ailamaki \cite{Kim2026Confining},
Sureshkumar \cite{Sureshkumar2026RLAM} &
Workflow architecture &
Explicit structures constrain the influence of probabilistic decisions on
execution. &
These approaches support architectural containment but do not define a general
criterion for the permissible effect range within the process. \\

Dhage \cite{Dhage2026Harness} &
Empirical constraint evaluation &
Finite-state control, forced tool selection, and output validation alone had
mixed effects on run-to-run reproducibility. Adding schema validation of plans
before tool invocation made most configurations fully reproducible. &
Shows that the effect of such control mechanisms can be measured, which
provides a basis for evaluating Envelope boundaries empirically. \\

Wang et al. \cite{Wang2023SelfConsistency} &
Aggregation across runs &
Multiple generated solution paths are consolidated into a common result. &
Can be used as a consolidation mechanism within a variation space before
exit. \\

Pattnayak and Bhatia \cite{Pattnayak2026ReproEvalCard},
Gu et al. \cite{Gu2025OLMES} &
Reproducibility &
Documented configurations and controlled conditions make repeated runs
comparable. &
Provides a basis for defining and documenting the controlled input state. \\

Rosen and Rosen \cite{Rosen2026Lineage},
Mudasiru \cite{Mudasiru2026Replay},
Khatchadourian \cite{Khatchadourian2026Replayable} &
Execution lineage and reconstruction &
Execution paths and external interactions are recorded so that they can later
be traced or reconstructed. &
Supports traceability of the execution that occurred within a variation
space. \\

\bottomrule
\end{tabularx}
\end{table*}

\subsection{Non-Determinism and Reproducibility of LLM Inference}
\label{sec:rw-inference}

Even with settings intended to be deterministic, repeated LLM calls may
produce different results. Atıl et al. demonstrate such run-to-run deviations
for hosted LLM systems, while Yuan et al. investigate numerical causes of
these deviations \cite{Atil2025NonDeterminism,Yuan2025Numerical}. Methods such
as LLM-42 and CoRun attempt to reduce causes of non-determinism already at the
inference level \cite{Gond2026LLM42,Zhao2026CoRun}.

From the process perspective considered here, it is not necessary to make
every LLM execution technically deterministic. What matters is whether
different executions produce differences in process-relevant properties. Technical
non-determinism must therefore be distinguished from non-deterministic process
behavior. A varying internal execution may remain irrelevant to the process if
its results are process-equivalent.

\subsection{Propagation in Agents and Multi-Step Systems}
\label{sec:rw-propagation}

In agentic systems, LLM outputs influence subsequent decisions, tool calls,
and states. Repeated runs may therefore develop different execution paths, or
\emph{trajectories}. Bjarnason et al. and Yagubyan show that such paths can
diverge at an early stage without necessarily leading to different final
results \cite{Bjarnason2026Randomness,Yagubyan2026Consistency}. Different
execution paths become relevant only when they lead to differences in process-relevant
properties.

Work on uncertainty and error propagation further shows that local deviations
can be propagated and amplified across multiple processing steps
\cite{Zhao2025SAUP,Xia2026Propagation,Mazhar2026Contamination,Xie2026Cascades,Singh2026Snowball}.
For the present approach, this raises the question of how far such differences
may extend within a process before control is required.

\subsection{Architectures for Constraining Probabilistic Effects}
\label{sec:rw-architecture}

Several approaches separate probabilistic processing from authoritative
system effects. The \emph{Stochastic-Deterministic Boundary} (SDB) describes
the transition from a probabilistically generated proposal to an
authoritative system action in terms of Proposer, Verifier, Commit, and Reject
\cite{Srinivasan2026SDB}. Schema-Gated Orchestration similarly separates
flexible LLM processing from execution controlled through machine-checkable
schemas \cite{Strickland2026Schema}.

The OWASP Agent Control Standard (ACS) makes this form of runtime control more
concrete. Standardized lifecycle hooks allow a Guardian to allow, deny,
modify, or escalate agent actions \cite{OWASP2026ACS,OWASP2026ACSRepo}.
In particular, they provide concrete technical intervention points before
tool calls.

APPA (\emph{Agentic Permissions Policy Algebra}) takes a more narrowly
information-flow-oriented approach. The LLM is treated as policy-untrusted. A
dual-phase reference monitor prospectively checks a tool call before execution
and validates the realized return again before it is admitted to the agent
context \cite{Kravchenko2026APPA}. For processing untrusted data, APPA can also
spawn temporary child branches whose context and label state remain separated
from the parent trajectory. The open-source implementation OpenAPPA, published
as a preview and request for comments, realizes these mechanisms as an agent-independent runtime and policy layer
\cite{OpenAPPA2026}. For the present work, APPA therefore provides a concrete
example of how individual control boundaries of a Deterministic Envelope can
be realized technically. APPA does not, however, determine in general where an
Envelope boundary should lie in a process or which differing results
are process-equivalent. SDB, schema gates, ACS, and APPA therefore address
different technical realizations of control boundaries, while the approach
proposed here derives their placement and extent from their process effects.

Other approaches constrain the variable region more strongly. Compiled AI
moves LLM processing into an upstream phase and then executes a validated
artifact deterministically \cite{Trooskens2026Compiled}. Workflow approaches
constrain probabilistic decisions through explicit process structures
\cite{Qiu2025Blueprint,Kim2026Confining,Sureshkumar2026RLAM}.

Dhage measures the effect of such constraints on repeated agent runs in
regulated tasks \cite{Dhage2026Harness}. Finite-state control, forced
tool selection, and output validation alone had mixed effects on
reproducibility. Only the additional schema validation of a plan before tool
invocation removed the remaining variance in most configurations, with
model-dependent effects on latency. This shows that the influence of control
mechanisms on process behavior can be evaluated experimentally. The higher-level
question of how far a region of permissible variation may extend within a
process remains open here as well.

\subsection{Reproducibility, Execution Lineage, and Reconstruction}
\label{sec:rw-replay}

For a process to be traceable, it must be possible to understand how an
authoritative state was produced. Research on reproducibility,
\emph{Execution Lineage}, and \emph{Deterministic Replay} shows how relevant
configurations, intermediate artifacts, dependencies, and external
interactions can be recorded or reconstructed
\cite{Pattnayak2026ReproEvalCard,Gu2025OLMES,Rosen2026Lineage,Mudasiru2026Replay,Khatchadourian2026Replayable}.
This does not require every internal LLM execution to be reproduced exactly.
Instead, the information needed to explain how an authoritative successor
state was derived from the preceding controlled state and why the transition
was accepted must be retained.

In multi-step workflows, variable processing may already cause effects outside
the controlled processing environment. In transactional systems, ACID
properties provide an established mechanism for committing changes atomically
or rolling them back after an abort \cite{Gray1981Transaction}. Distributed
workflows often lack such a general rollback mechanism. Saga-based approaches
address this problem through predefined compensation steps
\cite{GarciaMolina1987Sagas}. If an effect cannot be compensated or reversed,
it may therefore need to be controlled before it is executed.

\subsection{Research Gap and Research Questions}
\label{sec:research-gap}

The work reviewed above addresses important subproblems. Deterministic
inference techniques reduce the emergence of variation, studies of agentic
systems demonstrate its possible propagation, architectural approaches
provide control mechanisms for individual transitions, and work on execution
lineage supports later traceability. What remains open is the higher-level
process question of how far LLM-induced variation may extend within a
controlled process before another control point becomes necessary.

A control boundary immediately after every LLM call is not necessarily
required when different internal executions are equivalent with respect to
the downstream process. A later control point, however, may be too late if
differences have already created authoritative or irreversible effects. Such
a control boundary describes how a result is checked at a particular
transition. By itself, it does not determine where that transition should be
located within the process.

In the following, the region of permissible variation is referred to as the
\emph{controlled variation space}, and the corresponding architectural model
as the \emph{Deterministic Envelope}. Existing mechanisms such as SDB, schema
gates, ACS hooks, or transactional mechanisms can be employed at its
boundaries \cite{Srinivasan2026SDB,Strickland2026Schema,OWASP2026ACS,Gray1981Transaction}.
We investigate how the placement and extent of the variation space can be
determined and how results can be transferred into authoritative process
states in a controlled manner.

The study is guided by three research questions.

\noindent\textbf{RQ1.} Under what conditions does variation in an LLM-based
processing step become relevant to a controlled process?

\noindent\textbf{RQ2.} How can the effects of such variation be constrained
within a process so that different internal execution paths do not
uncontrollably lead to different authoritative process states?

\noindent\textbf{RQ3.} What information and mechanisms are required to trace
and verify the creation of an authoritative process state despite varying LLM
execution?

The questions build on one another. RQ1 distinguishes mere technical variation
from process-relevant deviation. RQ2 examines how its effects can be
constrained and where control points are required. RQ3 complements this
perspective by addressing the traceability of the transition into an
authoritative process state.

\section{Controlled Variation Space}
\label{sec:envelope}

\begin{figure*}[t]
\centering
\begin{adjustbox}{max width=\textwidth}
\begin{tikzpicture}[
  scale=1.0,
  every node/.style={transform shape},
  font=\small,
  box/.style={
    draw,
    rounded corners=2pt,
    minimum height=8mm,
    align=center,
    inner sep=4pt
  },
  state/.style={
    box,
    fill=white,
    minimum width=22mm
  },
  gate/.style={
    box,
    fill=tablehead,
    minimum width=16mm
  },
  traj/.style={
    -{Latex[length=2mm]},
    thick
  }
]

\node[state] (sin) at (0,0)
  {$s_i^{-}$\\[-2pt]
   {\scriptsize controlled}\\[-3pt]
   {\scriptsize input state}};

\coordinate (bl) at (2.9,-1.95);
\coordinate (br) at (8.9,1.60);

\begin{scope}[on background layer]
  \fill[envfill,rounded corners=3pt] (bl) rectangle (br);
\end{scope}

\draw[dashed,rounded corners=3pt] (bl) rectangle (br);

\node[anchor=south west,font=\scriptsize] at (2.9,1.68)
  {Variation space $B_i$};

\draw[traj]
  (sin.east)
  .. controls (3.4,0.55) and (5.2,0.90) ..
  (7.3,0.50)
  node[pos=0.72,above=1pt,font=\scriptsize]
  {$\tau_{B_i}^{(a)}$};

\draw[traj]
  (sin.east)
  .. controls (3.4,-0.75) and (5.2,-1.10) ..
  (7.3,-0.45)
  node[pos=0.72,below=1pt,font=\scriptsize]
  {$\tau_{B_i}^{(b)}$};

\node[font=\scriptsize,align=center] at (5.9,-0.05)
  {LLM calls, tool interactions,\\[-2pt]
   deterministic intermediate steps};

\node[gate] (g) at (10.3,0)
  {$G_i$\\[-2pt]
   {\scriptsize exit}\\[-3pt]
   {\scriptsize check}};

\node[state] (sout) at (13.3,0)
  {$s_i^{+}$\\[-2pt]
   {\scriptsize authoritative}\\[-3pt]
   {\scriptsize successor state}};

\draw[traj] (7.35,0.50) -- (g.north west);
\draw[traj] (7.35,-0.45) -- (g.south west);
\draw[traj] (g) -- (sout);

\draw[-{Latex[length=2mm]},dashed,gray]
  (g.south)
  .. controls (9.6,-2.45) and (5.6,-2.85) ..
  (4.6,-2.02)
  node[midway,below=5pt,fill=white,inner sep=1pt,font=\scriptsize,text=black]
  {not accepted, reprocessing or escalation};

\node[font=\scriptsize,anchor=north] at (6.6,-3.35)
  {Run artifact
   $R_i^{(r)}=
   \left(
   s_i^{-},
   \ \tau_{B_i}^{(r)},
   \ g_i^{(r)},
   \ s_i^{+}
   \right)$};

\end{tikzpicture}
\end{adjustbox}

\caption{Deterministic Envelope. Between the controlled input state and the
authoritative successor state, runs may develop different trajectories. The
exit check $G_i$ determines whether a result satisfies the exit conditions and
may become authoritative.}
\label{fig:envelope}
\end{figure*}
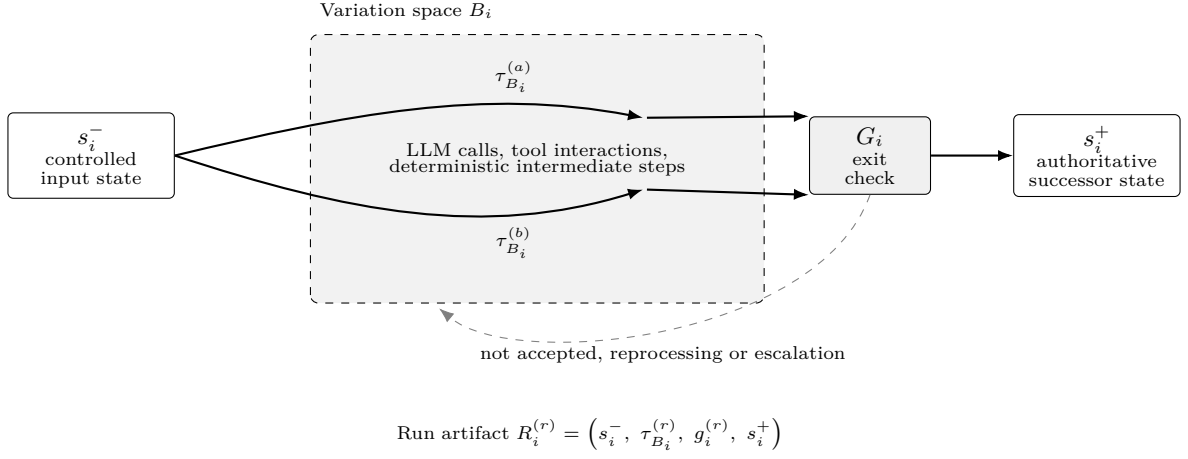

LLM-based processing can produce different results and execution paths from
the same initial state. We allow that variation deliberately, but only inside
a \emph{controlled variation space}. The region starts from a controlled input
state $s_i^{-}$ and ends at an exit check $G_i$. Until that check succeeds, LLM
outputs, tool results, and intermediate states remain working state rather than
authoritative process state. Only an accepted result is adopted as the
authoritative successor state $s_i^{+}$.

We use the term \emph{Deterministic Envelope} for this controlled transition
between authoritative states. The processing inside the Envelope may vary. The
rule that decides whether its result is adopted must not vary with it.

\subsection{Process-Relevant Equivalence}
\label{sec:local-process-effect}

We do not treat every difference between LLM outputs as a difference that
matters to the process. We care only about the properties that downstream
processing actually uses. Let $\pi_i$ denote a projection onto these
process-relevant properties. We call two results $y$ and $y'$ process-equivalent
if
\begin{equation}
y \equiv_i y'
\quad\Longleftrightarrow\quad
\pi_i(y) = \pi_i(y').
\label{eq:semanticvariation}
\end{equation}
For a coding agent, $\pi_i$ may include tested behavior, affected interfaces,
new dependencies, security findings, and modified paths. The comparison
therefore concerns properties that matter to the software process rather than
the concrete representation of the patch.

\subsection{Variation Space and State Transition}
\label{sec:variation-space}

The variation space $B_i$ may contain multiple LLM calls, tool interactions,
and deterministic processing steps. For a run $r$, $\tau_{B_i}^{(r)}$ denotes
the resulting execution path and $y_i^{(r)}$ its result. Different execution
paths are generally permitted within the variation space. The states produced
there have working-state character and are not yet authoritative states of the
process.

The basic structure of a Deterministic Envelope can be represented
schematically as
\begin{equation}
s_i^{-}
\longrightarrow
B_i
\longrightarrow
G_i
\longrightarrow
s_i^{+}.
\label{eq:state-transition-envelope}
\end{equation}
The controlled input state $s_i^{-}$ represents the accepted state from which
the LLM-based processing starts. Before execution, we define the
process-relevant properties $\pi_i$ and the exit conditions $C_i$. We also
record the task configuration, including the prompt, model version, inference
parameters, and, where applicable, tool configuration and permissions. These
elements remain fixed during execution.

A result is accepted if its process-relevant projection satisfies the exit
conditions
\begin{equation}
G_i\!\left(s_i^{-},\pi_i(y_i)\right)=\mathrm{accept}
\quad\Longleftrightarrow\quad
\pi_i(y_i)\models C_i.
\label{eq:exit-conditions}
\end{equation}
The conditions in $C_i$ may include tests, rules, schemas, policy checks, or
human approval. Only an accepted result may become the next authoritative
process state $s_i^{+}$. A rejected result remains outside the authoritative
process state.

Process-equivalent results have the same projection $\pi_i$ and are therefore
indistinguishable for the downstream process as defined by that projection.
This does not mean that only one result can be accepted. The exit conditions
may admit several different process-relevant projections. Another run may
therefore produce a different admissible result. This is a robustness question
rather than a failure of the controlled transition. Robustness can be increased
by narrowing the variation space, moving derivations into deterministic rules,
comparing several runs \cite{Wang2023SelfConsistency}, or adding human review
where the stakes are high.

\subsection{Chaining Multiple Deterministic Envelopes}
\label{sec:chain}

A process may contain multiple LLM-based sections. The accepted successor
state of one Envelope then becomes the controlled input state of the next.
This creates a sequence of authoritative checkpoints between which
non-deterministic processing can take place under control.

A checkpoint is an accepted process state whose creation has been validated
by the preceding exit check. Subsequent process steps may therefore use it as
an authoritative starting point. A checkpoint provides a stable reference
between two variation spaces and may be versioned or persisted for later
checks, reruns, or reconstruction of the process history. If a downstream
section must be executed again, it can resume from the most recently accepted
checkpoint without re-executing the preceding variable processing.

The accepted successor state $s_i^{+}$ becomes the controlled input state
$s_{i+1}^{-}$ of the next Envelope. The chaining can therefore be represented
schematically as
\begin{equation}
s_i^{-}
\to B_i
\to G_i
\to s_i^{+}=s_{i+1}^{-}
\to B_{i+1}
\to G_{i+1}
\to s_{i+1}^{+}.
\label{eq:envelope-chain}
\end{equation}

The states $s_i^{+}$ and $s_{i+1}^{+}$ form authoritative checkpoints. A
rejected result must not become the starting point of a subsequent process
section. This prevents unaccepted variation from propagating into later parts
of the process.

\begin{figure*}[t]
\centering
\begin{adjustbox}{max width=\textwidth}
\begin{tikzpicture}[
  scale=0.88,
  every node/.style={transform shape},
  font=\small,
  box/.style={
    draw,
    rounded corners=2pt,
    minimum height=8mm,
    align=center,
    inner sep=4pt
  },
  state/.style={
    box,
    fill=white,
    minimum width=20mm
  },
  gate/.style={
    box,
    fill=tablehead,
    minimum width=15mm
  },
  arrow/.style={
    -{Latex[length=2mm]},
    thick
  }
]

\node[state] (s0) at (0,0)
  {$s_i$\\[-2pt]
   {\scriptsize authoritative}\\[-3pt]
   {\scriptsize state}};

\coordinate (b1l) at (1.75,-1.45);
\coordinate (b1r) at (5.15,1.35);

\begin{scope}[on background layer]
  \fill[envfill,rounded corners=3pt] (b1l) rectangle (b1r);
\end{scope}

\draw[dashed,rounded corners=3pt]
  (b1l) rectangle (b1r);

\node[anchor=south west,font=\scriptsize] at (1.75,1.43)
  {Envelope $E_i$};

\node[font=\scriptsize,align=center] at (3.45,0.15)
  {$B_i$\\
   LLM, tools,\\
   working states};

\node[gate] (g1) at (6.25,0)
  {$G_i$\\[-2pt]
   {\scriptsize check}};

\node[state] (s1) at (8.45,0)
  {$s_{i+1}$\\[-2pt]
   {\scriptsize authoritative}\\[-3pt]
   {\scriptsize checkpoint}};

\coordinate (b2l) at (9.95,-1.45);
\coordinate (b2r) at (13.35,1.35);

\begin{scope}[on background layer]
  \fill[envfill,rounded corners=3pt] (b2l) rectangle (b2r);
\end{scope}

\draw[dashed,rounded corners=3pt]
  (b2l) rectangle (b2r);

\node[anchor=south west,font=\scriptsize] at (9.95,1.43)
  {Envelope $E_{i+1}$};

\node[font=\scriptsize,align=center] at (11.65,0.15)
  {$B_{i+1}$\\
   LLM, tools,\\
   working states};

\node[gate] (g2) at (14.45,0)
  {$G_{i+1}$\\[-2pt]
   {\scriptsize check}};

\node[state] (s2) at (16.75,0)
  {$s_{i+2}$\\[-2pt]
   {\scriptsize authoritative}\\[-3pt]
   {\scriptsize checkpoint}};

\draw[arrow] (s0.east) -- (1.75,0);
\draw[arrow] (5.15,0) -- (g1.west);
\draw[arrow] (g1.east) -- (s1.west);
\draw[arrow] (s1.east) -- (9.95,0);
\draw[arrow] (13.35,0) -- (g2.west);
\draw[arrow] (g2.east) -- (s2.west);

\draw[-{Latex[length=2mm]},dashed,gray]
  (g1.south)
  .. controls (6.0,-2.0) and (3.7,-2.15) ..
  (3.1,-1.48)
  node[midway,below=5pt,fill=white,inner sep=1pt,font=\scriptsize,text=black]
  {not accepted, defined handling};

\draw[-{Latex[length=2mm]},dashed,gray]
  (g2.south)
  .. controls (14.2,-2.0) and (11.9,-2.15) ..
  (11.3,-1.48)
  node[midway,below=5pt,fill=white,inner sep=1pt,font=\scriptsize,text=black]
  {not accepted, defined handling};

\node[font=\scriptsize,align=center] at (8.45,-2.85)
  {Authoritativeness is established at checkpoints.\\
   Between them, variation remains permitted under control.};

\end{tikzpicture}
\end{adjustbox}

\caption{Chaining multiple Deterministic Envelopes. A successfully checked
successor state forms an authoritative checkpoint and may serve as the
controlled input state of the next variation space. The checkpoint also
provides a stable reference for traceability and re-execution. If an exit
check fails, the result is not adopted as the next authoritative process
state.}
\label{fig:envelope-chain}
\end{figure*}
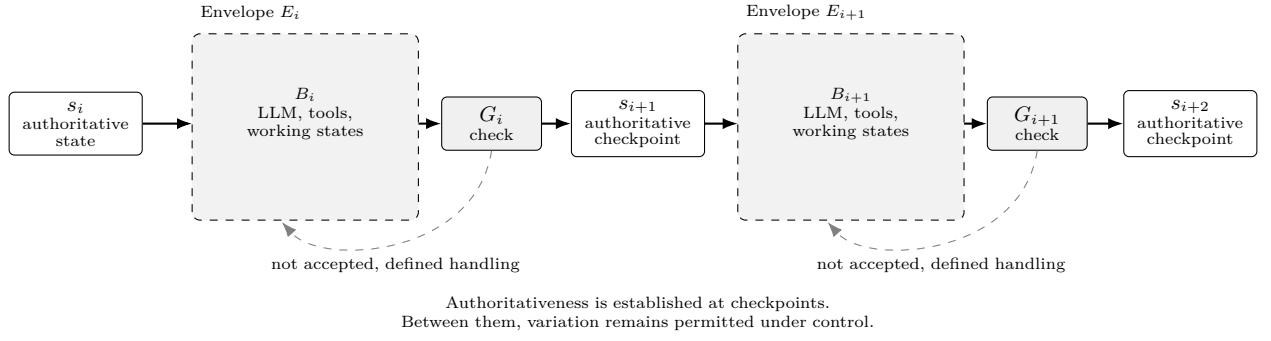

Figure~\ref{fig:envelope-chain} illustrates how multiple Envelopes can be
combined into a controlled process. The following section applies this model
to software development, where checkpoints may correspond, for example, to
accepted requirements, reviewed code changes, validated builds, or approved
releases.

\subsection{Placement of Boundaries}
\label{sec:congruence}

An Envelope boundary is required where variation can acquire authoritative
effect on the downstream process. This applies in particular to results that
serve as the basis for subsequent process steps, modify external states, or
have security- or regulatory-relevant effects. As long as different internal
executions remain process-equivalent, they may remain within the same
variation space. If differences can affect downstream processing, control is
required.

The position of the boundary therefore follows from the potential
\emph{process effect} rather than from the position of an LLM or tool call
alone. Technical mechanisms such as schema gates or ACS hooks can realize the
corresponding control points \cite{OWASP2026ACSRepo}.

\subsection{Exit Check}
\label{sec:exit-conditions}

The exit check $G_i$ in \eqref{eq:exit-conditions} determines whether the
result produced in the variation space may become an authoritative successor
state. It evaluates the process-relevant projection $\pi_i(y_i)$ against the
exit conditions $C_i$ that were fixed before execution.

$G_i$ does not denote a specific technical mechanism. Depending on the process
context, the exit conditions may be checked by schema validation, tests,
business rules, evidence requirements, policy checks, or human approval.
Several checks may also be combined.

Defining the exit conditions $C_i$ is itself a non-trivial design task. The
Deterministic Envelope does not assume that the complete desired result can be
specified in advance or expressed as an exhaustive set of tests. Such a
requirement would largely shift the problem toward full pre-specification and,
in software development, resemble a strict form of test-driven development.

Instead, $C_i$ defines the conditions under which a variable result may become
authoritative. These conditions may combine task-specific acceptance criteria
with controls that already exist in the surrounding process, such as regression
tests, interface checks, dependency policies, security rules, protected files,
and human approval. Only the conditions specific to the current transition
need to be added where existing controls are insufficient.

This distinction reduces specification effort but does not eliminate it. If a
process-relevant property is neither represented by $\pi_i$ nor covered by
$C_i$, the Envelope cannot ensure that different accepted results behave
equivalently with respect to that property. The quality of the controlled
transition therefore depends on the adequacy of the selected process-relevant
properties and exit conditions.

The required checks also depend on the significance of the subsequent state.
A local working-state change may require only lightweight validation, whereas
a production change, external write operation, or security-relevant action may
require stronger checks or explicit approval. The exit check is therefore the
point at which the variation permitted within the Envelope is bound again to
the requirements of the process.

\subsection{External Effects}
\label{sec:effects}

Effects may already arise within a variation space before the exit check, for
example through tool calls or modifications of external resources. Such
effects must be considered when defining Envelope boundaries. Within a
transactional system, a transition can be executed atomically. ACID
transactions provide an established basis for this purpose
\cite{Gray1981Transaction}. Changes become authoritative only after a
successful check or are rolled back on abort.

For distributed tool calls and external services, however, a common
transaction mechanism is often unavailable. Write operations or irreversible
actions must therefore either be controlled before execution, isolated, or
compensated where possible \cite{GarciaMolina1987Sagas}. The Deterministic
Envelope thus does not assume a general rollback capability. Instead, it
requires external effects to be permitted only where their consequences are
compatible with the control requirements of the process.

\subsection{Traceability}
\label{sec:traceability-replay}

For each run, the information required to trace how an accepted successor
state emerged from a controlled input state should be retained. The elements
of the Deterministic Envelope shown in Fig.~\ref{fig:envelope} are therefore
combined in a run artifact.

A \emph{run artifact} is the structured documentation of a concrete run. It
is not itself the process result. Instead, it records the essential
information required for later traceability of that result. This includes the
input state, the execution path produced within the variation space, the check
results relevant to the exit decision, and the finally accepted successor
state.

For an accepted run $r$, the run artifact can be represented abstractly as
\begin{equation}
R_i^{(r)} =
\left(
s_i^{-},
\tau_{B_i}^{(r)},
g_i^{(r)},
s_i^{+}
\right).
\label{eq:run-artifact}
\end{equation}
Here, $\tau_{B_i}^{(r)}$ denotes the execution path produced within the
variation space. The term $g_i^{(r)}$ contains the check results produced in
the concrete run on which the exit check $G_i$ based its decision. These may
include, for example, tool calls, modified files, test results, security
findings, or approval decisions.

The tuple above summarizes the core run-specific elements. The run artifact
also records the exit conditions and the task configuration under which the
transition was accepted. This includes the prompt, model version, inference
parameters, and, where applicable, tool configuration and permissions.

The run artifact thereby connects the controlled input state, the relevant
execution, the checking step, and the resulting authoritative successor state.
It can be stored as evidence of an individual run and used for later audits or
for reconstructing the process history. The objective is not to reproduce an LLM execution token by token. Instead,
it must be possible to trace the basis on which an authoritative successor
state was produced and why the exit conditions were considered satisfied.

\section{Application to Regulated Software Development}
\label{sec:applications}

We use regulated software development as the concrete application of the
Deterministic Envelope. The setting is useful because proposed changes can be
checked against requirements, tests, and policies, while intermediate work may
already touch files, dependencies, tools, or external systems. It therefore
makes the distinction between working state and authoritative software state
particularly visible.

In regulated software development, the authoritative state does not consist
of an individual code proposal produced by the LLM. Rather, it is the approved
software artifact together with its documented derivation. Starting from the
same initial state, a coding agent may choose different implementation paths.
Different patches are therefore not inherently problematic.

The controlled input state $s_i^{-}$ includes, in particular, the relevant
requirements version, the repository commit, and the applicable development
and quality rules. Within the variation space $B_i$, the agent may analyze
code, generate implementation variants, modify files, and add tests. These
working states are not yet authoritative process states.

The process-relevant projection $\pi_i$ focuses on the properties relevant to
approval. These may include requirements satisfaction, interface behavior,
permitted dependencies, security conditions, affected files, and other quality
properties. The exit check $G_i$ may combine builds, tests, static analysis,
security checks, and policy checks. In higher-risk cases, an additional review
may be required.

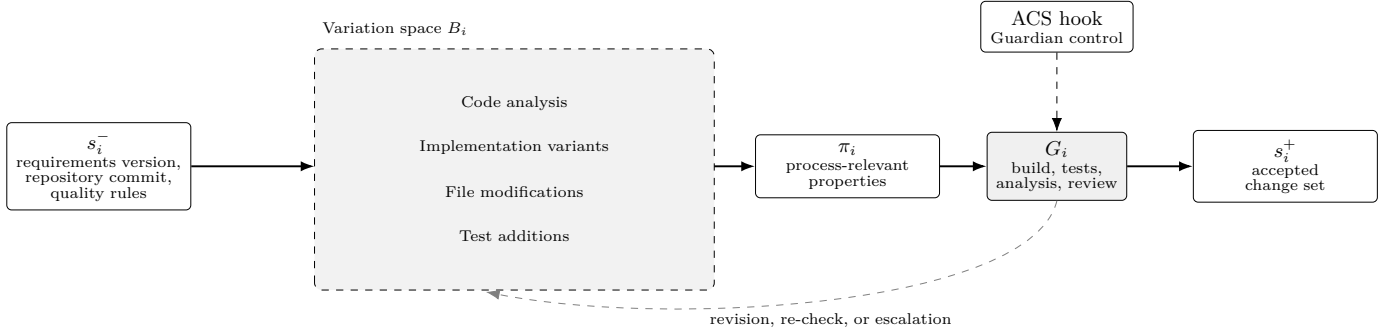
\begin{figure*}[t]
\centering
\begin{adjustbox}{max width=\textwidth}
\begin{tikzpicture}[
  scale=0.92,
  every node/.style={transform shape},
  font=\small,
  box/.style={
    draw,
    rounded corners=2pt,
    minimum height=8mm,
    align=center,
    inner sep=4pt
  },
  state/.style={
    box,
    fill=white,
    minimum width=29mm
  },
  gate/.style={
    box,
    fill=tablehead,
    minimum width=22mm
  },
  arrow/.style={
    -{Latex[length=2mm]},
    thick
  }
]

\node[state] (sin) at (0,0)
  {$s_i^{-}$\\[-2pt]
   {\scriptsize requirements version,}\\[-3pt]
   {\scriptsize repository commit,}\\[-3pt]
   {\scriptsize quality rules}};

\coordinate (bl) at (3.4,-1.95);
\coordinate (br) at (9.7,1.85);

\begin{scope}[on background layer]
  \fill[envfill,rounded corners=3pt]
    (bl) rectangle (br);
\end{scope}

\draw[dashed,rounded corners=3pt]
  (bl) rectangle (br);

\node[anchor=south west,font=\scriptsize]
  at (3.4,1.93)
  {Variation space $B_i$};

\node[font=\scriptsize,align=center]
  at (6.55,1.00)
  {Code analysis};

\node[font=\scriptsize,align=center]
  at (6.55,0.30)
  {Implementation variants};

\node[font=\scriptsize,align=center]
  at (6.55,-0.40)
  {File modifications};

\node[font=\scriptsize,align=center]
  at (6.55,-1.10)
  {Test additions};

\draw[arrow]
  (sin.east) -- (3.4,0);

\node[box,fill=white,minimum width=29mm] (proj)
  at (11.8,0)
  {$\pi_i$\\[-2pt]
   {\scriptsize process-relevant}\\[-3pt]
   {\scriptsize properties}};

\node[gate] (g)
  at (15.1,0)
  {$G_i$\\[-2pt]
   {\scriptsize build, tests,}\\[-3pt]
   {\scriptsize analysis, review}};

\node[state] (sout)
  at (18.7,0)
  {$s_i^{+}$\\[-2pt]
   {\scriptsize accepted}\\[-3pt]
   {\scriptsize change set}};

\draw[arrow]
  (9.7,0) -- (proj.west);

\draw[arrow]
  (proj.east) -- (g.west);

\draw[arrow]
  (g.east) -- (sout.west);

\node[box,fill=white,minimum width=24mm] (acs)
  at (15.1,2.2)
  {ACS hook\\[-2pt]
   {\scriptsize Guardian control}};

\draw[-{Latex[length=2mm]},dashed]
  (acs.south) -- (g.north);

\draw[-{Latex[length=2mm]},dashed,gray]
  (g.south)
  .. controls (14.7,-2.2) and (9.0,-2.55) ..
  (6.1,-1.98)
  node[midway,below=5pt,fill=white,inner sep=1pt,font=\scriptsize,text=black]
  {revision, re-check, or escalation};

\end{tikzpicture}
\end{adjustbox}

\caption{Application of the Deterministic Envelope to regulated software
development. The requirements baseline, repository commit, and quality rules
define the controlled input state. Within the variation space, analysis, code
generation, and modification steps may vary. Only a validated change set whose process-relevant properties satisfy the
exit conditions becomes authoritative. An ACS hook
may serve as a technical control point.}
\label{fig:swe-envelope}
\end{figure*}

Figure~\ref{fig:swe-envelope} maps the model introduced in
Section~\ref{sec:envelope} to regulated software development. The requirements
baseline, repository commit, and quality rules form the controlled input state.
Analysis, code generation, and modification steps take place within the
variation space. Before a result is adopted by the downstream process, the
properties relevant to approval are evaluated by the exit check.

A control of this kind can, for example, be implemented technically through
lifecycle hooks of the OWASP Agent Control Standard
\cite{OWASP2026ACS,OWASP2026ACSRepo}. Before a critical agent action, a
Guardian may check whether the conditions defined for the Envelope are
satisfied and whether the action may be executed. ACS thus provides one
possible technical intervention mechanism for a control boundary determined
at the process level. The position of this boundary, however, is not determined
by the hook itself, but by the potential effect of the agent action on the
process.

Only after the required checks have succeeded is the change set adopted as the
authoritative successor state $s_i^{+}$. The Deterministic Envelope therefore
does not require identical patches across repeated runs. It requires that a
varying result cannot cross the approval boundary without satisfying the
defined exit conditions.

The case also illustrates the distinction made in
Section~\ref{sec:envelope} between working state and authoritative process
state. Changes inside an isolated development environment may remain within
the variation space. A production commit, an approval, or another irreversible
effect, by contrast, requires prior control. For traceability, the input state,
the relevant execution artifacts, the check results, and the accepted
successor state are associated with the respective run.

\emph{Example.} Consider a coding agent that is to add IBAN validation to a
payment API. The controlled input state contains the accepted repository commit
and requirement REQ-142, which states that the payment endpoint must reject an
invalid IBAN with HTTP status code 422. The projection $\pi_i$ maps a change
set to the properties relevant for approval, including test results, interface
behavior, new dependencies, security findings, and affected paths. Differences
in naming, code structure, or comments are disregarded. The exit conditions
$C_i$ are listed in Table~\ref{tab:swe-conditions}. Before execution, the prompt
template, model version, inference parameters, and, where applicable, tool
configuration and permissions are also recorded.

The acceptance conditions must remain outside the agent's control. Write access
to the acceptance tests, CI configuration, and dependency allowlist is therefore
restricted. Condition $c_7$ additionally verifies that the acceptance tests
used for REQ-142 have not been modified. Most of these conditions need not be
defined specifically for REQ-142. Regression tests, dependency policies,
interface checks, static analysis, and protected file scopes may already be
part of the development process. Only requirement-specific acceptance behavior
has to be added where existing controls are insufficient.

Consider three runs from the same controlled input state. In the first run, the
agent implements the IBAN validation directly and satisfies $c_1$ to $c_7$.
Invalid IBANs are rejected with HTTP status code 422, the regression tests pass,
and no prohibited dependency or interface change is introduced. The change set
is accepted and becomes the authoritative successor state $s_i^{+}$.

In the second run, the agent uses an external IBAN-validation library that is
not contained in the approved dependency allowlist. Although the resulting
implementation may provide the required functional behavior, condition $c_3$
fails. The result is therefore rejected and cannot become authoritative.

In the third run, the agent implements the same required behavior using a
different code structure. It also satisfies all exit conditions and has the
same projection under $\pi_i$ as the first result. The first and third results
are therefore process-equivalent despite their different implementations.

The example also illustrates a limitation of the control definition. Suppose
REQ-142 and the acceptance tests do not specify whether lowercase IBANs should
be accepted. One implementation may accept them while another rejects them,
although both satisfy the current exit conditions $C_i$. In this case, the
relevant behavior has not yet been captured by $\pi_i$ and $C_i$. If this
behavior matters to the process, the requirement must be refined and an
additional acceptance condition introduced for subsequent transitions.

The run artifact introduced in Section~\ref{sec:traceability-replay} records
the controlled input state, task configuration, execution trace, check results,
and accepted successor state. An audit can therefore reconstruct why a
particular change became authoritative without requiring the original LLM
execution to be repeated.

\begin{table}[t]
\centering
\caption{Exit conditions $C_i$ for the IBAN-validation example}
\label{tab:swe-conditions}
\scriptsize
\renewcommand{\arraystretch}{1.15}

\begin{tabularx}{\columnwidth}{
@{}
>{\raggedright\arraybackslash}p{0.45cm}
>{\raggedright\arraybackslash}X
>{\raggedright\arraybackslash}p{2.35cm}
@{}
}
\toprule
 & \textbf{Condition} & \textbf{Check} \\
\midrule
$c_1$ &
Invalid IBANs are rejected with HTTP status code 422. &
Acceptance tests for REQ-142 \\

$c_2$ &
All existing regression tests pass. &
CI regression test suite \\

$c_3$ &
No new dependency outside the approved allowlist is introduced. &
Dependency changes checked against approved allowlist \\

$c_4$ &
The public API signature remains unchanged. &
Comparison of generated OpenAPI specification \\

$c_5$ &
No new static-analysis finding of medium or higher severity is introduced. &
SAST comparison with baseline \\

$c_6$ &
Changes are confined to \texttt{src/payments/}. &
Changed-file paths checked against allowed scope \\

$c_7$ &
The acceptance tests used for REQ-142 remain unchanged. &
Diff check of protected acceptance-test files \\

\bottomrule
\end{tabularx}
\end{table}

\section{Conceptual Results, Discussion, and Limitations}
\label{sec:fqanswers}

The following subsections bring together the related work, the model developed
in Section~\ref{sec:envelope}, and the software development application from
Section~\ref{sec:applications}. They examine the process relevance of
variation, the constraint of its effects, and the traceability of
authoritative state transitions.

\subsection{Process Relevance of Variation (RQ1)}
\label{sec:fq1}

A deviation is not process-relevant merely because two LLM outputs differ in
their wording, structure, or execution path. What matters are the properties
that are relevant to the downstream process. The process-relevant projection
$\pi_i$ maps a result to these properties. Different outputs may therefore be
equivalent for the downstream process even though their concrete representation or
derivation differs.

Variation becomes process-relevant when process-relevant properties differ and
that difference can influence downstream processing. In software development,
this may occur when two runs introduce different dependencies, change different
interfaces, or produce different security findings. Work on uncertainty
propagation and diverging agent trajectories indicates that local differences
can continue to affect subsequent processing steps
\cite{Zhao2025SAUP,Bjarnason2026Randomness,Xia2026Propagation}.

We define $\pi_i$ before the run so that what counts as relevant is not decided
after the result is known.

We therefore answer RQ1 in terms of process effect. LLM-induced variation
becomes relevant when it changes properties that matter to a subsequent
decision or authoritative process state. Technical, linguistic, or structural
differences remain irrelevant as long as the process-relevant projection treats them as
equivalent.

\subsection{Constraining the Effects of Variation (RQ2)}
\label{sec:fq2}

The Deterministic Envelope does not constrain variation by requiring every
individual LLM step to execute deterministically. Instead, it defines a
controlled variation space $B_i$ within which different execution paths are
permitted. A result is checked only at a defined exit boundary and is then,
where appropriate, adopted as an authoritative process state.

The position of this boundary is determined by the potential process effect.
As long as differences can be treated as internal working states and do not
acquire authoritative effect, multiple LLM and tool steps may remain within a
single Envelope. A boundary becomes necessary before a result is processed
further as authoritative or before an effect arises that can no longer be
controlled sufficiently.

Existing approaches can technically realize such boundaries. A
Stochastic-Deterministic Boundary structures the transition between proposal,
verification, and commit \cite{Srinivasan2026SDB}. Schema gating checks
generated artifacts before execution \cite{Strickland2026Schema}. ACS
provides standardized runtime hooks for this purpose
\cite{OWASP2026ACS,OWASP2026ACSRepo}. APPA complements these mechanisms with
prospective information-flow checks before tool calls, a second check when
return data is admitted to the agent context, and temporarily confined
processing paths for untrusted data \cite{Kravchenko2026APPA}. Compiled AI
moves the variable region entirely ahead of deterministic runtime for suitable
tasks \cite{Trooskens2026Compiled}.

External effects constrain the possible extent of an Envelope. Within a
transactional system, ACID properties can commit changes atomically and
isolate intermediate states from other processing
\cite{Gray1981Transaction}. For the case considered here, isolation is
particularly relevant. If intermediate results from a variable process section
that has not yet been accepted become visible or effective to other process
steps, this separation no longer holds. Downstream steps may then build on a
state that is later rejected or assessed differently. This can also violate
process-level consistency conditions of the overall process.

In distributed systems and external services, such transactional isolation is
often unavailable. In those settings, upstream controls or, where possible,
compensation mechanisms such as Sagas can be used
\cite{GarciaMolina1987Sagas}. Non-compensable or irreversible effects may
require control before execution. Multiple Envelopes can be chained into
longer processes, provided that only a successfully checked state becomes the
authoritative input to the next variation space.

The exit conditions $C_i$ are fixed before execution and must not be modified
within $B_i$. This prevents the variable processing from adapting the criteria
by which it is judged. The Envelope thus ensures that authoritative successor states arise only
from results accepted against the specified exit conditions and, through
checkpoints, remain stable for the completed transition. Robustness against
other admissible results of further runs requires a narrower variation space
or, where the stakes are high, redundancy.

Our answer to RQ2 is therefore not to constrain every internal step. We bound
the region in which variation may occur, fix the exit checks in advance, and
control the transition into an authoritative state. We place the boundary at
the point where internal variation can acquire authoritative or insufficiently
reversible effects on the downstream process.

\subsection{Traceability of Authoritative States (RQ3)}
\label{sec:fq3}

Traceability of an authoritative process state does not require identical
repetition of the internal LLM execution. Instead, the information explaining
which controlled input state produced the result, which process-relevant
artifacts and external interactions were involved, which exit check was
performed, and which authoritative successor state resulted must be retained.
These elements are brought together in the run artifact introduced in
Section~\ref{sec:traceability-replay}. The run artifact also records the exit
conditions and task configuration under which the state became authoritative.

For software development, the relevant information includes the requirements
version, repository state, task configuration, change set, tool interactions,
and check results. These elements must remain linked to the accepted software
state.

Approaches based on \emph{Execution Lineage} and \emph{Deterministic Replay}
provide technical foundations for this purpose
\cite{Rosen2026Lineage,Mudasiru2026Replay}. Execution lineage denotes the
traceable association of execution steps, intermediate artifacts,
dependencies, and external interactions with a concrete run. Deterministic
replay denotes techniques by which such an execution can be reconstructed or
re-executed in a controlled manner on the basis of recorded states and events.
The objective is not to reproduce the same token sequence or the complete
execution path (\emph{trajectory}), but to reconstruct the process-relevant
state transition and the associated acceptance decision.

For RQ3, we require a continuous link from the controlled input state through
the relevant execution and evidence artifacts to the exit decision and the
authoritative successor state. This is the information an auditor needs to
reconstruct why a particular result, rather than merely an LLM response, was
adopted by the process.

\subsection{Design Guidelines for Non-Deterministic Transitions}
\label{sec:guidelines}

The model implies that two design decisions precede every run. The first is
whether a software-development step needs non-deterministic processing at all.
The second is how the task is posed to the coding agent so that the exit check
$G_i$ can decide on acceptance. The following guidelines derive from
Sections~\ref{sec:envelope} and~\ref{sec:applications}.

\emph{G1. Use non-determinism only where it is necessary.} Before a step is
assigned to an LLM at runtime, it should be checked whether deterministic rules
or code suffice, or whether an LLM can be used at design time to produce a
validated artifact that then runs deterministically
\cite{Trooskens2026Compiled}. Only work that benefits from semantic interpretation, planning, or code
generation belongs in a runtime variation space.

\emph{G2. Prefer transitions whose results are easier to check than to
produce.} An autonomous LLM-based transition is appropriate primarily where
verifying a result is substantially cheaper than generating it, as with a code
change checked against existing acceptance and regression tests. If checking is as hard as producing, $G_i$ effectively reduces
to human judgment, and the LLM acts as an assistant rather than as an
autonomous process transition. Combined with the reversibility of the effect,
this yields the design space shown in Table~\ref{tab:design-matrix}.

\emph{G3. Fix the exit criteria before execution.} Acceptance tests, rules,
and thresholds form the exit conditions $C_i$. They are fixed before the run
and cannot be changed within the variation space. If the LLM generates or changes the
criteria by which its own result is judged, the control boundary loses its
independence. This guideline shares APPA's stance of
treating the LLM as untrusted for control decisions
\cite{Kravchenko2026APPA}, but transfers it from security policies on
information flows to the process-level exit conditions of a process
transition. How a task must be posed so that its result becomes checkable is
not addressed by information-flow control.

\emph{G4. Make the checked properties explicit.} The agent should expose the
properties captured by $\pi_i$ in a form that can be checked directly. In
software development, these may include test results, affected interfaces,
dependency changes, security findings, and modified paths. $G_i$ should not
have to infer these properties from an unconstrained narrative. Schema validation before execution has been shown to remove remaining
run-to-run variance in constrained agent workflows
\cite{Strickland2026Schema,Dhage2026Harness}.

\emph{G5. Require independently verifiable evidence.} A proposed change should
be accompanied by evidence that can be generated or checked independently of
the model, such as build and test results, static-analysis findings, documented
dependency changes, and the set of affected files or paths. This separates the
agent's proposal from the evidence used to justify its acceptance and reduces
reliance on the model's own assessment of its output.

\emph{G6. Allow explicit escalation.} If a requirement is ambiguous, a needed
permission is unavailable, or the agent cannot produce sufficient evidence, it
should be allowed to stop and request review. Forcing a result encourages the
agent to guess and weakens the value of the exit check.

\emph{G7. Version the task.} Record the prompt, model version, inference
parameters, and, where applicable, tool configuration and permissions in the
run artifact \eqref{eq:run-artifact}. Without this information, the execution
conditions under which an authoritative state was produced cannot be
reconstructed completely.

These guidelines do not guarantee that an accepted result is correct. They
move residual uncertainty to points where it becomes visible and can be
escalated, and they reduce the share of a transition that depends on
non-deterministic judgment. Their effect has not yet been evaluated
empirically.

\begin{table}[t]
\centering
\caption{Design space for LLM-based transitions by checkability and reversibility of the effect}
\label{tab:design-matrix}
\scriptsize
\renewcommand{\arraystretch}{1.15}
\begin{tabularx}{\columnwidth}{
@{}
>{\raggedright\arraybackslash}p{1.6cm}
>{\raggedright\arraybackslash}X
>{\raggedright\arraybackslash}X
@{}
}
\toprule
 & \textbf{Easy to check} & \textbf{Hard to check} \\
\midrule
\textbf{Reversible} &
Wide Envelope with automated acceptance by $G_i$ &
Acceptance with sampling or subsequent review \\
\textbf{Irreversible} &
Narrow Envelope with a strict check before the effect &
Human approval, or no LLM at runtime \\
\bottomrule
\end{tabularx}
\end{table}

\subsection{Discussion}
\label{sec:discussion}

Our results move the determinism requirement away from complete reproducibility
of model execution and toward control of its process effects. This gives the
LLM room to vary where semantic flexibility is useful, but it also exposes the
price of that freedom. We must be able to state which properties matter to
the process and when a state transition becomes authoritative.

The process-relevant projection $\pi_i$ determines which differences are considered
relevant to downstream processing. If it is too coarse, relevant differences
may be hidden. If it is too fine-grained, even irrelevant differences become
significant and the benefit of the variation space decreases. The software-development
example considered here admits a plausible projection. A general method for
deriving such projections from requirements, however, has not yet been
provided.

The exit check also does not provide an absolute guarantee. Its validity is
limited to the properties that are actually checked. A test suite may, for
example, assess functional behavior without covering every security or
architectural property. A schema may validate the structure of an artifact
without ensuring its semantic correctness. The same limitation applies to
existing gate, boundary, and runtime-control approaches
\cite{Strickland2026Schema,Srinivasan2026SDB,OWASP2026ACS}. APPA illustrates
this limitation particularly clearly. Its formal guarantees concern declared
information flows, registered policies, and the associated Trusted Computing
Base. They do not establish whether a content-level permissible result is
semantically correct, and confining a child trajectory does not roll back external
side effects that have already occurred \cite{Kravchenko2026APPA}. APPA is
therefore complementary to the Deterministic Envelope. It can strongly enforce
specific security-relevant boundaries, but it does not replace the process-level
definition of the variation space and its exit conditions.

The width of an Envelope is therefore a design decision subject to risk
trade-offs. Narrow boundaries limit the reach of potential deviations but
introduce more checks and transitions. Wide variation spaces preserve greater
semantic freedom but require stronger assumptions and controls at their
boundaries. Write operations and irreversible external actions in particular
limit how far an Envelope can extend in practice.

Deterministic inference remains a complementary strategy. Methods such as
LLM-42 and CoRun can reduce technical causes of variation
\cite{Gond2026LLM42,Zhao2026CoRun}. They do not, however, eliminate variation
arising from external tool results, changing information spaces, or
semantically different intermediate states. The Deterministic Envelope
therefore does not aim to eliminate variation completely, but to control its
effects on authoritative process states.

\subsection{Limitations and Validity}
\label{sec:limitations}

We deliberately keep the proposal conceptual, and several limitations follow
from that choice. Most importantly, the process-relevant projection $\pi_i$ has to be
defined for the specific process context. The model is only as good as that projection.
If it omits a property that later matters to a decision or authoritative state,
the Envelope cannot recover the omission. We do not yet provide a general
method for deriving such projections from requirements.

A further limitation concerns the exit check. The Deterministic Envelope can
assure only those properties that are actually covered by the checks employed.
An accepted transition therefore does not automatically imply that the result
is correct or risk-free in every respect. The quality of the resulting
process depends substantially on the completeness and validity of the tests,
rules, policies, and, where applicable, human reviews used.

External effects cannot be handled in a fully general manner either. While
changes within transactional systems may be withheld or rolled back,
corresponding mechanisms are not necessarily available for external services
or irreversible actions. Such effects may therefore require additional
control points before execution and may constrain the permissible extent of an
Envelope.

The model ensures admissibility with respect to the specified exit
conditions and stability of accepted states for the completed transition, but
not robustness across repeated runs. Another run may produce a different
admissible result
whenever the exit conditions admit several different process-relevant projections. The
software development example shows that such gaps usually stem from incomplete
specifications. The design guidelines in Section~\ref{sec:guidelines} are
derived conceptually and have not yet been evaluated empirically.

The software-development example is a conceptual illustration. It does not
constitute empirical validation and does not yet support robust conclusions
about transferability to other application areas or development settings. Measurements exist for individual
constraint mechanisms \cite{Dhage2026Harness}, but not for the placement of
Envelope boundaries. It therefore remains open whether, and to what extent,
Deterministic Envelopes improve the stability or traceability of authoritative
process states compared with uncontrolled LLM integration.

Empirical validation should therefore investigate how reliably process-relevant
equivalence criteria can be defined, how different placements of Envelope
boundaries affect resulting process states, and what additional control and
documentation overhead is introduced. The main validity risks concern the
operationalization of process-relevant equivalence and the transferability of results
that have so far been developed conceptually and illustrated through one
software-development example.

\subsection{Role as a Foundation for Concrete Control Architectures}
\label{sec:foundation}

We intentionally formulate Controlled Non-Determinism and the Deterministic
Envelope independently of a specific application context. We do not prescribe regulatory
roles or concrete software components. Instead, we use the model for the
upstream design decision. We ask where probabilistic processing should be
allowed to vary and where it must be rebound to a controlled process state. A concrete
control architecture can then realize that boundary through policies,
permissions, verification, approvals, or persistence mechanisms.

ACS and APPA illustrate the relationship between a conceptual boundary and a
technical control architecture. ACS provides standardized intervention points
at which agent actions can be controlled by a Guardian
\cite{OWASP2026ACS,OWASP2026ACSRepo}. For information-flow-critical tool
interactions, APPA additionally specifies a prospective check before execution
and a second check of the realized return before its admission to the agent
context \cite{Kravchenko2026APPA}. Such a hook or reference-monitor check can
technically implement a checkpoint or exit control. Likewise, the temporary
confinement of a sub-trajectory used by APPA can technically isolate a subregion
of a variation space. Not every technically available control point must,
however, constitute an Envelope boundary, and a boundary required by the
process may consist of several technical controls. The choice is determined by
the potential process effect of the variation. The Deterministic Envelope
describes \emph{where} control is required and which process effect is to be
bounded. ACS, APPA, and comparable mechanisms support \emph{how} parts of this
control can be implemented technically
\cite{OWASP2026ACS,OWASP2026ACSRepo,Kravchenko2026APPA}.

\subsection{Boundary Placement in Production Agent Environments}
\label{sec:environments}

Environments for agentic software development already implement such
boundaries, and they place them differently. The following comparison
reflects the vendor documentation as of October 2026. Claude Code intercepts individual
actions. Permission modes determine which classes of action run without a
prompt, rules and lifecycle hooks refine this per tool, and a separate sandbox
limits what an action can reach once it runs
\cite{Anthropic2026Permissions,Liu2026DesignSpace}. OpenAI Codex separates the
same two concerns explicitly. A sandbox mode defines the technical reach of a
step, while an approval policy determines which actions require confirmation
before execution \cite{OpenAI2026Codex}. Gemini CLI combines approval modes
with an optional sandbox and makes the trust status of a workspace a
precondition for autonomous execution. In an untrusted folder, tool execution
requires individual confirmation and project configuration is not loaded
\cite{Google2026GeminiCLI,Google2026GeminiTrust}.

OpenAPPA, by contrast, is not a coding agent of its own but an agent-independent
policy and runtime layer. It sits between an agent and its tools and checks
before a tool call whether the intended information flow and prior execution
history satisfy the configured rules. Return data can be checked again before
being admitted to the agent context. For untrusted data, the underlying APPA
model additionally supports confined child branches, provided that the harness
or proxy in use supplies the context separation required for them
\cite{Kravchenko2026APPA,OpenAPPA2026}. The control boundary therefore does not
reside in a single agent implementation but at the interface between agent,
tool, and context. Such a layer complements operating-system sandboxing rather
than replacing it, because trajectory confinement does not revert external
effects that have already been committed \cite{Kravchenko2026APPA}.

The GitHub Copilot cloud agent places an important control boundary further
downstream. It works in an isolated branch with restricted network access and
cannot approve or merge its own pull request. Pull-request review therefore
forms an important exit check \cite{GitHub2026CopilotAgent}.

Some of these environments additionally define classes of actions that cannot
be approved automatically. Codex requires approval for tool calls marked as
destructive, while Claude Code excludes selected critical removal operations
from automatic approval
\cite{OpenAI2026Codex,Anthropic2026Permissions}. This corresponds to the
requirement in Section~\ref{sec:effects} that non-compensable effects be
controlled before execution.

Two distinctions follow from the comparison, and
Table~\ref{tab:environments} summarizes them. The first concerns how the
variation space is bounded. Interception checks individual actions, whereas
isolation permits a broader section of work while constraining its technical
reach. The second concerns how the boundary is enforced or evaluated. Static
rules and technical isolation mechanisms can provide deterministic controls.
Human review and model-based checks, in contrast, introduce an additional
variable decision component.

\begin{table}[t]
\centering
\caption{Placement of control boundaries in environments for agentic software development}
\label{tab:environments}
\scriptsize
\renewcommand{\arraystretch}{1.15}
\begin{tabularx}{\columnwidth}{
@{}
>{\raggedright\arraybackslash}p{1.75cm}
>{\raggedright\arraybackslash}X
>{\raggedright\arraybackslash}p{2.85cm}
@{}
}
\toprule
\textbf{Environment} &
\textbf{Bounding of the variation space} &
\textbf{Check performed by} \\
\midrule
Claude Code \cite{Anthropic2026Permissions} &
Interception per action, with a sandbox as an additional layer &
Human, rule, hook, or classifier model \\
OpenAI Codex \cite{OpenAI2026Codex} &
Sandbox reach, with approval on escalation &
Human or policy \\
Gemini CLI \cite{Google2026GeminiCLI,Google2026GeminiTrust} &
Optional sandbox, with workspace trust as a precondition &
Human or approval mode \\
OpenAPPA / APPA \cite{Kravchenko2026APPA,OpenAPPA2026} &
Reference monitor before tool execution and context admission, optional
trajectory confinement &
Deterministic policy, registered authority where applicable \\
Copilot cloud agent \cite{GitHub2026CopilotAgent} &
Isolated branch with restricted network access &
Pull-request review \\
\bottomrule
\end{tabularx}
\end{table}

\section{Conclusion and Future Work}
\label{sec:conclusion}

LLM-based systems introduce variation into processes that are otherwise expected
to produce controlled and accountable state transitions. This paper has examined
how such variation can be accommodated without requiring every internal LLM
execution to be deterministic. Our main result is a process-oriented model that
confines non-deterministic processing to explicitly bounded regions and controls
where its results may acquire authoritative effect.

Our central claim is that deterministic processes do not require
identical internal LLM executions. They require a clear definition of which
variation is permissible, where that variation can acquire authoritative
effect, and under which conditions a result may become the next accepted
process state.

We capture this distinction with \emph{Controlled Non-Determinism} and the
\emph{Deterministic Envelope}. Inside a controlled variation space, LLMs,
tools, and deterministic processing steps may follow different execution
paths. Authoritativeness is established only at an explicit exit boundary.
Process-relevant equivalence separates differences that matter to the process
from those that are merely linguistic, structural, or technical.

Before execution, the relevant properties and exit conditions are defined and
the task configuration is recorded. These elements remain fixed while execution
takes place within the variation space. Only a result that satisfies the exit
conditions may become the next authoritative process state. Different runs may
still produce different admissible results. Robustness can be increased by
narrowing the variation space and moving suitable derivations into deterministic
rules. The design guidelines developed in this paper translate these principles
into requirements for posing LLM tasks, with a preference for transitions whose
results are easier to check than to produce.

Variation becomes relevant when it changes properties that affect subsequent
decisions or authoritative process states. Its effects can therefore be
constrained by encapsulating variable processing between controlled states and
checking results at appropriate boundaries. Traceability does not require
complete technical reproducibility. Instead, it requires a documented link
between the input state, relevant execution artifacts, the exit decision, and
the authoritative successor state.

Existing mechanisms such as Stochastic-Deterministic Boundaries, schema gates,
ACS hooks, or transactional techniques can technically realize such
boundaries \cite{Srinivasan2026SDB,Strickland2026Schema,OWASP2026ACS,Gray1981Transaction}.
The Deterministic Envelope addresses the upstream design question of
\emph{where} these boundaries are required within a process and how far
variable processing may extend before another control boundary is required.

The software-development example illustrates that the useful width of an
Envelope depends on potential process effect rather than on the number or
position of LLM calls. Analysis and code generation can remain inside a broader
variation space when their effects are isolated. Write operations,
security-relevant actions, and irreversible external effects require earlier or
stronger control. Boundary placement is therefore a process-design decision,
not a property of an individual model invocation.

A broader implication is that increasing automation of implementation may shift
software-engineering effort from producing code toward specifying and verifying
acceptable behavior. Requirements, tests, policies, contracts, and invariants
then become not only documentation or quality-assurance artifacts, but part of
the control structure that determines which generated results may become
authoritative. This gives renewed importance to formal and semi-formal
specification techniques, including ideas developed in early work on abstract
data types, algebraic specifications, axiomatic program reasoning, and precise
module specifications
\cite{Liskov1974ADT,GuttagHorning1978Algebraic,Hoare1969Axiomatic,Parnas1972Specification}.
The implementation may vary while the properties that an acceptable
implementation must preserve are stated as precisely as required by the
process. The Deterministic Envelope does not require a complete formal
specification, but its effectiveness depends on how well the relevant
properties and exit conditions capture the behavior that matters.

The present paper remains conceptual. We see three priorities for empirical
work. The first is to derive process-relevant equivalence criteria reliably
from requirements. The second is to measure how different boundary placements
affect the stability of authoritative states and the resulting control
overhead. The third is to test whether the proposed design guidelines improve
real software-development processes. A related question is how much
specification effort is required for different levels of control and to what
extent existing tests, policies, contracts, and other development artifacts
can be reused as exit conditions.

The environments compared in Section~\ref{sec:environments} provide a concrete
starting point for such validation. Their default control boundaries can be
described using the model developed here, making alternative placements
comparable. Existing studies of these environments have examined configuration
artifacts and internal design decisions rather than boundary placement
\cite{Galster2026Harness,Liu2026DesignSpace}. The effects of different
placements on state stability and control overhead could be evaluated using
methods similar to those applied to individual constraint mechanisms
\cite{Dhage2026Harness}.

One property of these environments also points beyond the current model. Where
an approval decision is delegated to a second model rather than to a rule or a
human reviewer \cite{Anthropic2026Permissions}, the exit check itself becomes
probabilistic. Under which conditions such a probabilistic check can still
reliably bound the propagation of variation remains an open question.

\section*{Acknowledgment}

Large language models were used during the preparation of this manuscript for
language editing, consistency checks, and suggestions concerning wording and
structure. The
research questions and the conceptual framework were developed by the authors,
who take full responsibility for the scientific content.

\FloatBarrier


\begin{thebibliography}{99}

\bibitem{Yuan2025Numerical}
Yuan, J., Li, H., Ding, X., Xie, W., Li, Y.-J., Zhao, W., Wan, K.,
Shi, J., Hu, X., Liu, Z.
\emph{Understanding and Mitigating Numerical Sources of Nondeterminism in LLM Inference}.
In: Advances in Neural Information Processing Systems 38 (NeurIPS 2025), 2025.
\url{https://proceedings.neurips.cc/paper_files/paper/2025/hash/f80094a824ba5912d4a2de169c404a40-Abstract-Conference.html}.

\bibitem{Gond2026LLM42}
Gond, R., Kamath, A.~K., Ramjee, R., Panwar, A.
\emph{LLM-42: Enabling Determinism in LLM Inference with Verified Speculation}.
arXiv:2601.17768, 2026.
doi: \url{https://doi.org/10.48550/arXiv.2601.17768}.

\bibitem{Zhao2026CoRun}
Zhao, S., Yang, J., Chen, Q., Hu, J., Zheng, J., Chen, G., Chen, X.
\emph{CoRun: Padding is Simple and Efficient for Deterministic LLM Inference}.
arXiv:2608.14376, 2026.
\url{https://arxiv.org/abs/2608.14376}.

\bibitem{Atil2025NonDeterminism}
Atıl, B., Aykent, S., Chittams, A., Fu, L., Passonneau, R.~J.,
Radcliffe, E., Rajagopal, G.~R., Sloan, A., Tudrej, T., Ture, F.,
Wu, Z., Xu, L., Baldwin, B.
\emph{Non-Determinism of ``Deterministic'' LLM System Settings in Hosted Environments}.
In: Proceedings of the 5th Workshop on Evaluation and Comparison of NLP Systems,
pp.~135--148, Association for Computational Linguistics, 2025.
doi: \url{https://doi.org/10.18653/v1/2025.eval4nlp-1.12}.

\bibitem{Zhao2025SAUP}
Zhao, Q., Li, D., Liu, Y., Cheng, W., Sun, Y., Oishi, M., Osaki, T.,
Matsuda, K., Yao, H., Zhao, C., Chen, H., Zhao, X.
\emph{Uncertainty Propagation on LLM Agent}.
In: Proceedings of the 63rd Annual Meeting of the Association for Computational Linguistics,
Volume 1: Long Papers, pp.~6064--6073, 2025.
doi: \url{https://doi.org/10.18653/v1/2025.acl-long.302}.

\bibitem{Xia2026Propagation}
Xia, B., Zhu, L., Gao, E., Lu, Q., Xue, M., Sejdinovic, D.
\emph{Uncertainty Propagation in LLM-Based Systems}.
arXiv:2604.23505, 2026.
\url{https://arxiv.org/abs/2604.23505}.

\bibitem{Bjarnason2026Randomness}
Bjarnason, B.~H., Silva, A., Monperrus, M.
\emph{On Randomness in Agentic Evals}.
ICLR 2026 Workshop on Agents in the Wild. arXiv:2602.07150, 2026.
\url{https://arxiv.org/abs/2602.07150}.

\bibitem{Yagubyan2026Consistency}
Yagubyan, A.
\emph{How Consistent Are LLM Agents? Measuring Behavioral Reproducibility in Multi-Step Tool-Calling Pipelines}.
arXiv:2605.28840, 2026.
\url{https://arxiv.org/abs/2605.28840}.

\bibitem{Mazhar2026Contamination}
Mazhar, A., Suri, H., Galhotra, S.
\emph{Trace-Level Analysis of Information Contamination in Multi-Agent Systems}.
In: Proceedings of the ACM Conference on AI and Agentic Systems (ACM CAIS '26), 2026.
doi: \url{https://doi.org/10.1145/3786335.3813147}.
Preprint: arXiv:2604.27586.

\bibitem{Xie2026Cascades}
Xie, Y., Zhu, C., Zhang, X., Zhu, T., Ye, D., Qi, M., Chen, H., Zhou, W.
\emph{From Spark to Fire: Modeling and Mitigating Error Cascades in LLM-Based Multi-Agent Collaboration}.
arXiv:2603.04474, 2026.
\url{https://arxiv.org/abs/2603.04474}.

\bibitem{Singh2026Snowball}
Singh, P., Pawar, B.
\emph{The Hallucination Snowball: Modeling Error Propagation as State Transitions in Multi-Agent LLM Pipelines}.
FAGEN Workshop (Failure Modes in Agentic AI), ICML 2026. arXiv:2608.14588, 2026.
\url{https://arxiv.org/abs/2608.14588}.

\bibitem{Srinivasan2026SDB}
Srinivasan, V.
\emph{A Methodology for Selecting and Composing Runtime Architecture Patterns for Production LLM Agents}.
arXiv:2605.20173, 2026.
\url{https://arxiv.org/abs/2605.20173}.

\bibitem{Strickland2026Schema}
Strickland, J., Vijeta, A., Moores, C., Bodek, O., Nenchev, B.,
Whitehead, T., Phillips, C., Tassenberg, K., Conduit, G., Pellegrini, B.
\emph{Talk Freely, Execute Strictly: Schema-Gated Agentic AI for Flexible and Reproducible Scientific Workflows}.
arXiv:2603.06394, 2026.
\url{https://arxiv.org/abs/2603.06394}.

\bibitem{OWASP2026ACS}
OWASP GenAI Security Project.
\emph{Agent Control Standard (ACS)}.
OWASP Resource, September 1, 2026.
\url{https://genai.owasp.org/resource/agent-control-standard-acs/}.
Accessed: October 3, 2026.

\bibitem{OWASP2026ACSRepo}
OWASP GenAI Security Project.
\emph{Agent Control Standard: The Open Standard for Runtime Agent Control}.
Version 0.1 (Public Preview), GitHub repository and specification, 2026.
\url{https://github.com/GenAI-Security-Project/agent-control-standard}.
Accessed: October 3, 2026.

\bibitem{Kravchenko2026APPA}
Kravchenko, A., Liventsev, V., Konstantinov, I., Iskhakov, I., Kukuy, M.
\emph{APPA: Recoverable Information-Flow Control for Real-World LLM Agents}.
arXiv:2607.24625v2, 2026.
doi: \url{https://doi.org/10.48550/arXiv.2607.24625}.

\bibitem{Trooskens2026Compiled}
Trooskens, G., Karlsberg, A., Sharma, A., De Brouwer, L., Van Puyvelde, M.,
Young, M., Thickstun, J., Alterovitz, G., De Brouwer, W.~A.
\emph{Compiled AI: Deterministic Code Generation for LLM-Based Workflow Automation}.
arXiv:2604.05150, 2026.
\url{https://arxiv.org/abs/2604.05150}.

\bibitem{Qiu2025Blueprint}
Qiu, L., Ye, Y., Gao, Z., Zou, X., Chen, J., Gui, Z., Huang, W.,
Xue, X., Qiu, W., Zhao, K.
\emph{Blueprint First, Model Second: A Framework for Deterministic LLM Workflow}.
arXiv:2508.02721, 2025.
doi: \url{https://doi.org/10.48550/arXiv.2508.02721}.

\bibitem{Kim2026Confining}
Kim, K., Ailamaki, A.
\emph{Confining Nondeterminism: AI-Driven Research Systems as DBMSs for Reliable, Non-Wasteful, Transparent, and Collaborative Research [Vision]}.
arXiv:2607.10508, 2026.
\url{https://arxiv.org/abs/2607.10508}.

\bibitem{Sureshkumar2026RLAM}
Sureshkumar, S.
\emph{R-LAM: Reproducibility-Constrained Large Action Models for Scientific Workflow Automation}.
arXiv:2601.09749, 2026.
\url{https://arxiv.org/abs/2601.09749}.

\bibitem{Dhage2026Harness}
Dhage, S.
\emph{Harness Engineering for Predictable Agentic Systems: An Empirical Study of Deterministic Execution Constraints}.
arXiv:2608.26197, 2026.
\url{https://arxiv.org/abs/2608.26197}.

\bibitem{Wang2023SelfConsistency}
Wang, X., Wei, J., Schuurmans, D., Le, Q.~V., Chi, E.~H., Narang, S.,
Chowdhery, A., Zhou, D.
\emph{Self-Consistency Improves Chain of Thought Reasoning in Language Models}.
In: International Conference on Learning Representations (ICLR), 2023.
doi: \url{https://doi.org/10.48550/arXiv.2203.11171}.

\bibitem{Pattnayak2026ReproEvalCard}
Pattnayak, P., Bhatia, A.
\emph{ReproEvalCard: A Reporting Standard for Reproducible Evaluation of LLM Pipelines}.
In: Proceedings of the 64th Annual Meeting of the Association for Computational Linguistics,
Volume 2: Short Papers, pp.~238--249, 2026.
doi: \url{https://doi.org/10.18653/v1/2026.acl-short.22}.

\bibitem{Gu2025OLMES}
Gu, Y., Tafjord, O., Kuehl, B., Haddad, D., Dodge, J., Hajishirzi, H.
\emph{OLMES: A Standard for Language Model Evaluations}.
In: Findings of the Association for Computational Linguistics: NAACL 2025,
pp.~5020--5048, 2025.
doi: \url{https://doi.org/10.18653/v1/2025.findings-naacl.282}.

\bibitem{Rosen2026Lineage}
Rosen, J., Rosen, S.
\emph{From Agent Loops to Deterministic Graphs: Execution Lineage for Reproducible AI-Native Work}.
arXiv:2605.06365, 2026.
\url{https://arxiv.org/abs/2605.06365}.

\bibitem{Mudasiru2026Replay}
Mudasiru, R.
\emph{Deterministic Replay for AI Agent Systems}.
arXiv:2607.16200, 2026.
\url{https://arxiv.org/abs/2607.16200}.

\bibitem{Khatchadourian2026Replayable}
Khatchadourian, R.
\emph{Replayable Financial Agents: A Determinism-Faithfulness Assurance Harness for Tool-Using LLM Agents}.
ICLR 2026 Workshop on Advances in Financial AI. arXiv:2601.15322, 2026.
\url{https://arxiv.org/abs/2601.15322}.

\bibitem{OpenAPPA2026}
Archestra AI.
\emph{OpenAPPA: Deterministic Guardrails That Don't Break Agents}.
Open-source implementation (preview/RFC), GitHub repository, 2026.
\url{https://github.com/archestra-ai/OpenAPPA}.
Accessed: October 3, 2026.

\bibitem{Gray1981Transaction}
Gray, J.
\emph{The Transaction Concept. Virtues and Limitations}.
In: Proceedings of the 7th International Conference on Very Large Data
Bases, pp.~144--154, 1981.

\bibitem{GarciaMolina1987Sagas}
Garcia-Molina, H., Salem, K.
\emph{Sagas}.
In: Proceedings of the 1987 ACM SIGMOD International Conference on
Management of Data, pp.~249--259, 1987.
doi: \url{https://doi.org/10.1145/38713.38742}.

\bibitem{Anthropic2026Permissions}
Anthropic.
\emph{Claude Code Documentation. Choose a Permission Mode}.
\url{https://code.claude.com/docs/en/permission-modes}.
Accessed: October 3, 2026.

\bibitem{Liu2026DesignSpace}
Liu, J., Zhao, X., Shang, X., Shen, Z.
\emph{Dive into Claude Code: The Design Space of Today's and Future AI Agent Systems}.
arXiv:2604.14228, 2026.
\url{https://arxiv.org/abs/2604.14228}.

\bibitem{OpenAI2026Codex}
OpenAI.
\emph{Codex Documentation. Agent Approvals and Security}.
\url{https://developers.openai.com/codex/agent-approvals-security}.
Accessed: October 3, 2026.

\bibitem{Google2026GeminiCLI}
Google.
\emph{Gemini CLI Documentation. Configuration Reference}.
GitHub repository, 2026.
\url{https://github.com/google-gemini/gemini-cli/blob/main/docs/reference/configuration.md}.
Accessed: October 3, 2026.

\bibitem{Google2026GeminiTrust}
Google.
\emph{Gemini CLI Documentation. Trusted Folders}.
GitHub repository, 2026.
\url{https://github.com/google-gemini/gemini-cli/blob/main/docs/cli/trusted-folders.md}.
Accessed: October 3, 2026.

\bibitem{GitHub2026CopilotAgent}
GitHub.
\emph{Risks and Mitigations for GitHub Copilot Cloud Agent}.
GitHub Docs, 2026.
\url{https://docs.github.com/en/copilot/concepts/agents/cloud-agent/risks-and-mitigations}.
Accessed: October 3, 2026.

\bibitem{Galster2026Harness}
Galster, M., Mohsenimofidi, S., Lulla, J.~L., Abubakar, M.~A., Treude, C.,
Baltes, S.
\emph{Harness Engineering for Agentic AI Coding Tools: An Exploratory Study}.
arXiv:2602.14690, 2026.
\url{https://arxiv.org/abs/2602.14690}.


\bibitem{Liskov1974ADT}
B.~H. Liskov and S.~N. Zilles,
``Programming with Abstract Data Types,''
in \emph{Proceedings of the ACM SIGPLAN Symposium on Very High Level Languages},
Santa Monica, CA, USA, 1974, pp.~50--59.
doi: \url{https://doi.org/10.1145/800233.807045}.

\bibitem{GuttagHorning1978Algebraic}
J.~V. Guttag and J.~J. Horning,
``The Algebraic Specification of Abstract Data Types,''
\emph{Acta Informatica},
vol.~10, no.~1, pp.~27--52, 1978.
doi: \url{https://doi.org/10.1007/BF00260922}.

\bibitem{Hoare1969Axiomatic}
C.~A.~R. Hoare,
``An Axiomatic Basis for Computer Programming,''
\emph{Communications of the ACM},
vol.~12, no.~10, pp.~576--580, Oct.~1969.
doi: \url{https://doi.org/10.1145/363235.363259}.

\bibitem{Parnas1972Specification}
D.~L. Parnas,
``A Technique for Software Module Specification with Examples,''
\emph{Communications of the ACM},
vol.~15, no.~5, pp.~330--336, May~1972.
doi: \url{https://doi.org/10.1145/355602.361309}.

\end{thebibliography}
\end{document}